\documentstyle[12pt,a4]{article}
\title{ Relativistic Hartree-Bogoliubov
theory with finite range pairing 
forces in coordinate space: Neutron halo in 
light nuclei}

\author{W. P\"oschl, D. Vretenar, 
G.A. Lalazissis and P. Ring\\
Physics Department, Technische Universit\"at  M\"unchen, D-85748 Garching}
\begin{document}
\maketitle
%#####################################################
% Abstract
%
%\maketitle
\begin{abstract}
%
%######################################################################
The Relativistic Hartree Bogoliubov (RHB) model is applied in the 
self-consistent mean-field approximation to the description of 
the neutron halo in the mass region above the s-d shell. 
Pairing correlations and the coupling to particle continuum states 
are described by finite range two-body forces.
Finite element methods are used 
in the coordinate space discretization of the coupled system
of Dirac-Hartree-Bogoliubov integro-differential eigenvalue equations,
and Klein-Gordon equations for the meson fields. Calculations are
performed for the isotopic chains of Ne and C nuclei. We find evidence 
for the occurrence of neutron halo in heavier Ne isotopes. The 
properties of the 1f-2p orbitals near the Fermi level and the 
neutron pairing interaction play a crucial role in the formation of the halo. 
Our calculations display no evidence for the neutron halo phenomenon in C 
isotopes.
\end{abstract}

%#####################################################################
\vskip 1.5cm 
\noindent 
The structure of exotic nuclei with extreme isospin 
values displays many interesting phenomena:
extremely weak binding of the outermost 
nucleons, coupling between bound states and the particle continuum, 
regions of neutron halos,  
large spatial dimensions and the existence of the neutron skin.
Modifications of shell structure in dripline nuclei
have been predicted, as well as changes in the evolution of collectivity. 
For dripline nuclei the separation 
energy of the last nucleons can become extremely small.
The Fermi level is found close to the 
particle continuum, and the lowest particle-hole or particle-particle 
modes are often embedded in the continuum. 
The neutron density 
distribution in such loosely bound nuclei shows an extremely long 
tail, the neutron halo. The resulting large interaction 
cross sections have provided the first experimental evidence 
for halo nuclei~\cite{Tani.85}. The neutron halo phenomenon
has been studied with a variety of theoretical models~\cite{Tani.95,HJJ.95}. 
For very light nuclei in particular, models based on the 
separation into core plus valence space nucleons 
(three-body Borromean systems) have been employed. 
In heavier neutron-rich nuclei one expects that 
mean-field models should 
provide a better description of ground-state properties.
In a mean-field description, the neutron halo and 
the stability against nucleon emission can only be explained 
with the inclusion of pairing correlations.
Both the properties of single-particle states near the 
neutron Fermi level and the pairing interaction are important
in the formation of the neutron halo. 

Relativistic mean-field models have been successfully applied in 
calculations of nuclear matter and properties of finite nuclei throughout 
the periodic table. The theory provides a framework for 
describing the nuclear many-body problem as a relativistic 
system of baryons and mesons.. 
In the self-consistent mean-field 
approximation, detailed calculations have been performed for a variety
of nuclear structure phenomena~\cite{Rin.96}.
The Relativistic Hartree-Bogoliubov (RHB) theory in coordinate
space, which is an extension of non-relativistic
HFB-theory~\cite{Doba.84}, provides a unified description of mean-field and
pairing correlations. 
The RHB theory has recently been applied in the description
of ground-state properties of Sn and Pb isotopes~\cite{Gonz.96}, using 
an expansion in a large oscillator basis for the 
solution of the Dirac-Hartree-Bogoliubov equations.
In many applications an expansion of the wave functions in an 
appropriate oscillator basis of spherical or axial symmetry provides
a satisfactory level of accuracy.  
For dripline nuclei, the expansion in the localized
oscillator basis presents only a poor approximation to 
the continuum states. 
Oscillator expansions display a slow 
convergence in the asymptotic region of the
coordinate space.
In order to correctly
describe the coupling between bound and 
continuum states, the Dirac-Hartree-Bogoliubov equations have to be 
solved in coordinate space. Recently, a fully self-consistent RHB model 
in coordinate space has been used to describe the two-neutron halo in 
$^{11}$Li~\cite{Meng.96}. However, only a density dependent 
force of zero range has been used in the pairing channel. 
Finite range forces, as for example the Gogny interaction, should 
provide a more realistic description of pairing correlations. 
A very efficient procedure for the coordinate space discretization  
of the RHB model is provided by Finite Element Methods (FEM).
In Refs.~\cite{PVR1.97,PVR2.97} we have developed a FEM based 
model and computer code 
for the solution of Dirac-Hartree-Bogoliubov equations in coordinate space.
In the present work we present the first application of the model in the
description of ground state properties of Ne and C isotopes. 
For these elements, isotopes near the neutron drip should soon become
accessible in experiments with radioactive beams. Therefore it is 
extremely important to investigate for which isotopes the 
relativistic Hartree-Bogoliubov model predicts the formation of 
neutron halo.
 
The model
describes the nucleus as
a system of Dirac nucleons which interact in a relativistic covariant manner
through the exchange of virtual mesons~\cite{Ser.92}:
the isoscalar scalar $\sigma$-meson,
the isoscalar vector $\omega$-meson and the isovector vector $\rho$-meson.
The photon field ~$(A)$ accounts for the electromagnetic interaction. 
The relativistic extension of the HFB theory is described in Ref.~ 
\cite{Kuch.91}. Independent quasi-particles are introduced and the ground state 
of a nucleus $\vert \Phi >$ is represented as the vacuum with 
respect to these quasi-particles. The quasi-particle operators 
are defined by a unitary Bogoliubov transformation of the 
single-nucleon creation and annihilation operators.
The generalized single-particle hamiltonian of HFB theory
contains two average potentials: the self-consistent field 
$\hat\Gamma$ which encloses all the long range {\it ph} correlations,
and a pairing field $\hat\Delta$ which sums up the 
{\it pp}-correlations. 
In the Hartree approximation for the self-consistent 
mean field, the Relativistic Hartree-Bogoliubov (RHB) equations read
\begin{eqnarray}
\left( \matrix{ \hat h_D -m- \lambda & \hat\Delta \cr
                -\hat\Delta^* & -\hat h_D + m +\lambda
                 } \right) \left( \matrix{ U_k \cr V_k } \right) =
E_k\left( \matrix{ U_k \cr V_k } \right).
\end{eqnarray}
where $\hat h_D$ is the single-nucleon Dirac hamiltonian, 
and $m$ is the nucleon mass.
$U_k$ and $V_k$ are  quasi-particle Dirac spinors, and $E_k$ denote 
the quasi-particle energies.

The RHB equations are non-linear 
integro-differential equations. They have to be solved self-consistently, 
with potentials determined in the mean-field approximation from 
solutions of Klein-Gordon equations for mesons
and Coulomb field:
\begin{eqnarray}
\bigl[-\Delta + m_{\sigma}^2\bigr]\,\sigma({\bf r})&=&
-g_{\sigma}\,
\sum\limits_{E_k > 0} V_k^{\dagger}({\bf r})\gamma^0 V_k({\bf r})
\nonumber \\
&\,&-g_2\,\sigma^2({\bf r})-g_3\,\sigma^3({\bf r}),   \\
\bigl[-\Delta + m_{\omega}^2\bigr]\,\omega^0({\bf r})&=&
~g_{\omega}\,
\sum\limits_{E_k > 0} V_k^{\dagger}({\bf r}) V_k({\bf r}), \\
\bigl[-\Delta + m_{\rho}^2\bigr]\,\rho^0({\bf r})&=&
~g_{\rho}\,
\sum\limits_{E_k > 0} V_k^{\dagger}({\bf r})\tau_3 V_k({\bf r}), \\
-\Delta \, A^0({\bf r})&=&
e\,\sum\limits_{E_k > 0} V_k^{\dagger}({\bf r}) {{1-\tau_3}\over 2} 
V_k({\bf r}).
\end{eqnarray}
The source terms are sums of bilinear 
products of baryon amplitudes. The sums run over all positive energy states.
The system of equations
is solved self-consistently in coordinate space
by discretization on the finite element mesh. 
In the coordinate space representation of the pairing field 
$\hat\Delta $, the kernel of the integral operator is
\begin{equation}
\Delta_{ab} ({\bf r}, {\bf r}') = {1\over 2}\sum\limits_{c,d}
V_{abcd}({\bf r},{\bf r}') {\bf\kappa}_{cd}({\bf r},{\bf r}').
\end{equation}
where 
$V_{abcd}({\bf r},{\bf r}')$ are matrix elements of a general 
two-body pairing interaction and 
${\bf\kappa}_{cd}({\bf r},{\bf r}')$, is 
the pairing tensor, defined as
\begin{equation}
\label{equ.4}
{\bf\kappa}_{cd}({\bf r},{\bf r}') := 
\sum_{E_k>0} U_{ck}^*({\bf r})V_{dk}({\bf r}').
\end{equation}
The integral operator $\hat\Delta$ acts on the wave function
$V_k({\bf r})$:
\begin{equation}
\label{equ.2.4}
(\hat\Delta V_k)({\bf r}) 
= \sum_b \int d^3r' \Delta_{ab} ({\bf r},{\bf r}') V_{bk}({\bf r}'). 
\end{equation}
In Ref.~\cite{Kuch.91} it has been argued that in principle one should use 
a one-meson exchange interaction $V_{abcd}$ in the 
pairing channel, as it is derived by the elimination of the mesonic 
degrees of freedom in the Lagrangian (1). However, it was also shown
that the standard parameter sets of the mean-field approximation 
lead to completely unrealistic pairing matrix elements.
These parameters do not reproduce the scattering data. Since at present
there exists no microscopic derivation of an appropriate
pairing interaction, we follow the prescription of Ref. \cite{Gonz.96},  
and use a phenomenological Gogny-type finite range interaction in the 
$pp$-channel (i.e. in Eq. 8), a procedure which requires no 
cut-off and which provides a very reliable description 
of pairing properties in finite nuclei.
\begin{eqnarray}
V^{pp}(1,2)~=~\sum_{i=1,2}
e^{-( ({\bf r}_1- {\bf r}_2) 
 / {\mu_i} )^2}\,
(W_i~+~B_i P^\sigma \nonumber \\
-H_i P^\tau -
M_i P^\sigma P^\tau),
\end{eqnarray}
with the parameters 
$\mu_i$, $W_i$, $B_i$, $H_i$ and $M_i$ $(i=1,2)$.

In what follows we present results of calculation 
for the even-even Ne and C isotopes. For the mean-field 
Lagrangian the NL3 parametrization has been used. This new 
parameter set has been derived recently~\cite{LKR.96}
by fitting ground state properties of a large number 
of spherical nuclei.
Properties calculated with the NL3 effective interaction 
are found to be in very good agreement with
experimental data for nuclei at and 
away from the line of beta-stability. The
parameter set D1S~\cite{BGG.84} has been used for 
the finite range Gogny type interaction.

In Fig. 1 the $rms$ radii for Ne (a) and C (b) isotopes
are plotted as functions of neutron number. We display 
neutron, proton and matter $rms$ radii, and the N$^{1/3}$
curves. These two curves are normalized so that they coincide 
with neutron $rms$ radii for $^{20}$Ne and $^{12}$C, respectively.
Neutron radii of Ne isotopes 
follow the N$^{1/3}$ curve up to N $\approx$ 22. 
For larger values of N a sharp increase of neutron radii 
is observed. The matter radii follow this boost, that is, 
by increasing the number of neutrons, the spatial extension of
the nucleus is substantially enlarged. In comparison, 
proton $rms$ radii are almost constant, 
and only display a slow constant increase. The last bound
isotope is $^{40}$Ne. For N $\geq$ 32 the neutron Fermi level
becomes positive. The sudden increase in neutron 
$rms$ radii, which is an indication for the halo
phenomenon, is not observed for the C isotopes. Both 
the neutron and matter radii display a monotonous increase
with the neutron number. 
The large neutron skin is a common
feature in neutron-rich $\beta$-unstable nuclei.
The last bound isotope is $^{24}$C,
i.e. we find bound systems only in the s-d shell. For N $>$ 20, 
in the region where for Ne the formation of the neutron halo is 
observed, the C isotopes are not bound.

For heavier Ne isotopes (N$\geq$ 20), 
in Fig. 2 we plot the proton and neutron density distributions. 
The proton density profiles do not change with the number 
of neutrons. The neutron density distributions display
an abrupt change between $^{30}$Ne and $^{32}$Ne. A long tail
emerges, revealing the formation of a multi-particle halo.
It should be noted, however, that the present calculations have been 
performed with the assumption of 
spherical symmetry, while several Ne isotopes are expected 
to be deformed. Therefore, what we observe is rather the trend of 
variations of $rms$ radii with addition of pairs of neutrons. 
An RHB code with which deformed nuclei could be calculated, is not
yet available. Nevertheless, we do not expect that the inclusion of 
deformation would alter the observed trend, especially for heavier
Ne isotopes.

In order to understand better the formation of neutron halo in 
the Ne isotopes, in Fig.3 we display the neutron single-particle
spectrum in the canonical basis. The energies of levels in the 
continuum decrease with increasing neutron number. The shell 
structure dramatically changes at the neutron drip N$\geq$ 22.
The triplet of states 1f$_{7/2}$, 2p$_{3/2}$ and  2p$_{1/2}$ 
approaches zero energy, and a gap is formed between these 
states and all other states in the continuum. The Fermi level 
uniformly increases toward zero for N$\leq$ 22. Between 
N = 22 and N = 32, the Fermi level is practically constant 
and very close to the continuum. The 
addition of neutrons in this region of the drip 
does not increase the binding. Only the 
spatial extension of neutron distribution displays an increase. 
At N = 32 the Fermi energy becomes 
slightly positive, and heavier isotopes are not bound any more.
The formation of the neutron halo is closely related to the
quasi-degeneracy of the triplet of states 1f$_{7/2}$, 2p$_{3/2}$ 
and  2p$_{1/2}$. The role of the 1f$_{7/2}$ is very interesting. 
The finite range pairing interaction promotes neutrons from the
1f$_{7/2}$ orbital to the 2p levels. 
Since these levels are 
so close in energy, the total binding energy does not change
significantly. Due to their small centrifugal barrier,
the 2p$_{3/2}$ and 2p$_{1/2}$ orbitals form the halo.
This is illustrated in Table 1 where
we display the occupation probabilities for the triplet of states.

The details are also shown in Fig. 4a, where only
the energies of these three levels and the Fermi energy are 
plotted as functions of the neutron number. The corresponding 
diagram for the $C$ isotopes is displayed in Fig. 4b. Here the
energy spacings between the triplet of states are much larger, 
and the levels are higher in the continuum. In particular, the 
1f$_{7/2}$ level is much higher in $C$ nuclei. For N = 20, with the 
s-d shell fully occupied, all three levels are still high in the
continuum. The pairing interaction does not have 
enough strength to promote particles in these levels, and
as a result, heavier isotopes are unbound. Therefore,
an interesting result of our calculation is that the RHB model, 
with finite range pairing interaction, predicts the formation of 
multi-neutron halo in $Ne$ isotopes with two protons above 
the Z=8 shell closure, but no halo is found for $C$ isotopes 
with two protons below the closed shell.

To illustrate the importance of the pairing interaction in the 
formation of the neutron halo, in Fig. 5 we display 
the radial dependence 
of the self consistent pairing field $\Delta$(r) for three Ne nuclei. 
For N $\leq$ 20, the contribution of the pairing interaction 
is relatively small. We find a sharp increase in pairing between
$^{30}$Ne and $^{32}$Ne. The 
pairing field is concentrated 
on the surface of the nucleus, the matter radius increases and
we observe the formation of the neutron halo. Of course the
pairing field increases further with the number of neutrons, but 
on a much smaller scale.

In conclusion, we report results of the first 
application of the relativistic Hartree Bogoliubov
model with finite range pairing interaction in
coordinate space. 
In particular, we have investigated 
the formation of neutron halo in dripline
Ne and C nuclei. The pairing interaction has a 
unique role in dripline nuclei, 
due to the scattering of neutron pairs 
from bound to continuum states. It is 
therefore of considerable interest to 
study the predictions of a finite range 
pairing force. 
We have found
evidence for the formation of multi-particle 
neutron halo in heavier Ne isotopes, but not in 
C nuclei. Our calculations show that, in these 
relatively
light nuclei, the halo phenomenon is caused 
by a delicate interplay of the self-consistent
mean field, which includes all the long 
range $ph$ correlations, and the pairing 
field, which sums up the $pp$ correlations and
describes the coupling to the continuum.
In Ne, the triplet of almost degenerate 
states 1f$_{7/2}$, 2p$_{3/2}$
and  2p$_{1/2}$, approaches zero energy at N = 20. 
The pairing interaction promotes neutrons from the 
1f to the 2p orbitals, and a  halo is formed. In
C on the other hand, the triplet of states
is still high in the continuum at the neutron
dripline, and the pairing force does not have enough 
strength to form a halo. The model describes 
the halo as a purely microscopic phenomenon.
We have also performed calculations for O isotopes,
and the results are similar to those for C. At 
N = 20 the 1f$_{7/2}$ level is still at $\approx$ 4 MeV
in the continuum, and $^{28}$O is the last bound isotope 
for the NL3 parameter set. The results for N, O, F, Na 
and Mg will be included in a forthcoming study on the 
influence of the proton number on the formation of the 
neutron halo in the mass region above the s-d shell. 

The authors thank D. Habs and G. M\"unzenberg for stimulating discussions.
This work has been supported by the 
Bundesministerium f\"ur Bildung und Forschung under 
project 06 TM 875. D. Vretenar is Alexander von Humboldt Fellow,
on leave of absence from University of Zagreb, Croatia.
\
%\section*{References}

\newpage
%%%%%%%%%%%%%%%%%%%%%%%%%%%%%%%%%%%%%%%%%%%%%%%%%%%%%%
\begin{table}
\caption{1f-2p occupation probabilities for the 
$Ne$ isotopes. }
\begin{center}
\begin{tabular}{llll}
N  & 1f$_{7/2}$ & 1p$_{3/2}$ & 1p$_{1/2}$ \\
\hline  
22 & 0.174 & 0.130 & 0.021 \\
24 & 0.291 & 0.350 & 0.064 \\
26 & 0.388 & 0.577 & 0.167 \\
28 & 0.480 & 0.749 & 0.411 \\
30 & 0.580 & 0.852 & 0.741 \\
32 & 0.709 & 0.929 & 0.924 \\
\end{tabular}
\end{center}
\end{table}
%%%%%%%%%%%%%%%%%%%%%%%%%%%%%%%%%%%%%%%%%%%%%%%%%%%%%%%%%%%%%

\centerline{\bf Figure Captions}

\begin{itemize}
\item{\bf Fig.1} Calculated $rms$ radii for Ne (a), and C (b) isotopes 
as functions of neutron number.

\item{\bf Fig.2} Proton and neutron density distribution for Ne isotopes.

\item{\bf Fig.3} Canonical basis single-particle neutron levels
as functions of the number of neutrons.The spectrum
is calculated for Ne isotopes.

\item{\bf Fig.4} 1f-2p single-particle neutron levels in the canonical 
basis, for the Ne (a), and C (b) isotopes.

\item{\bf Fig.5} Self consistent pairing field
for $^{30,32,40}$Ne

\end{itemize}

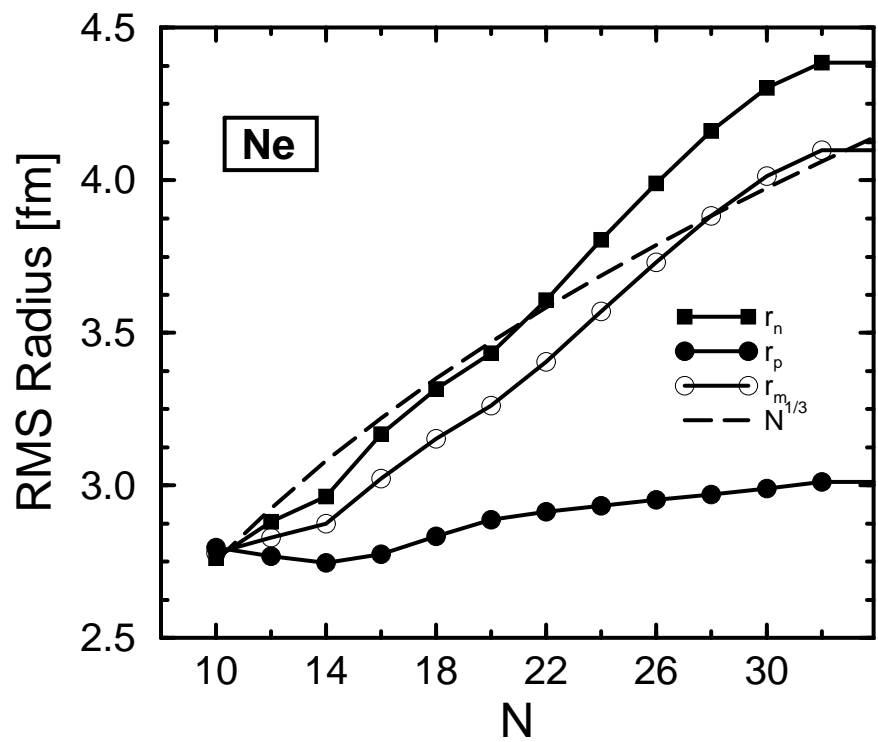
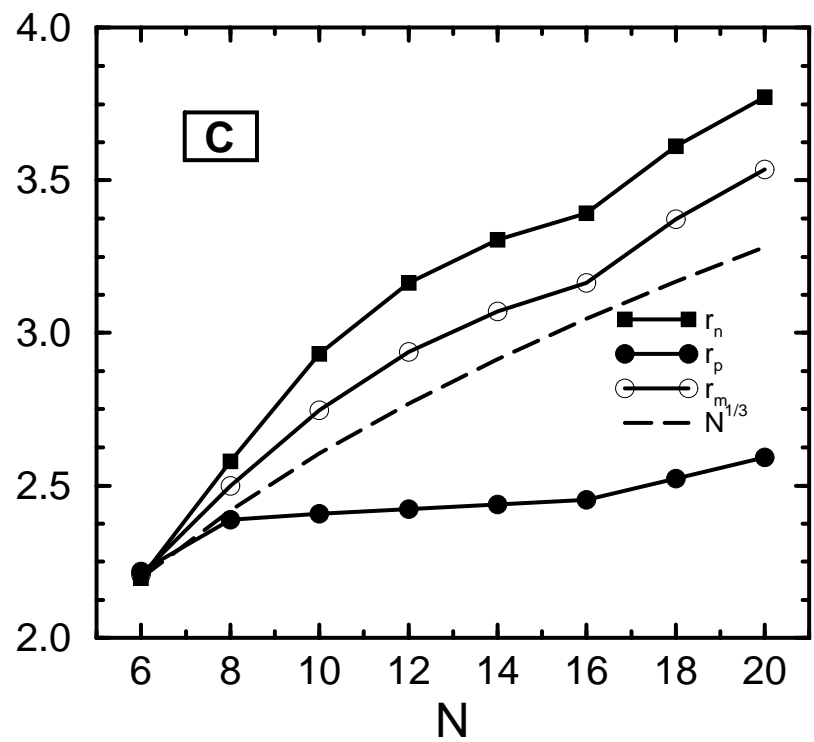
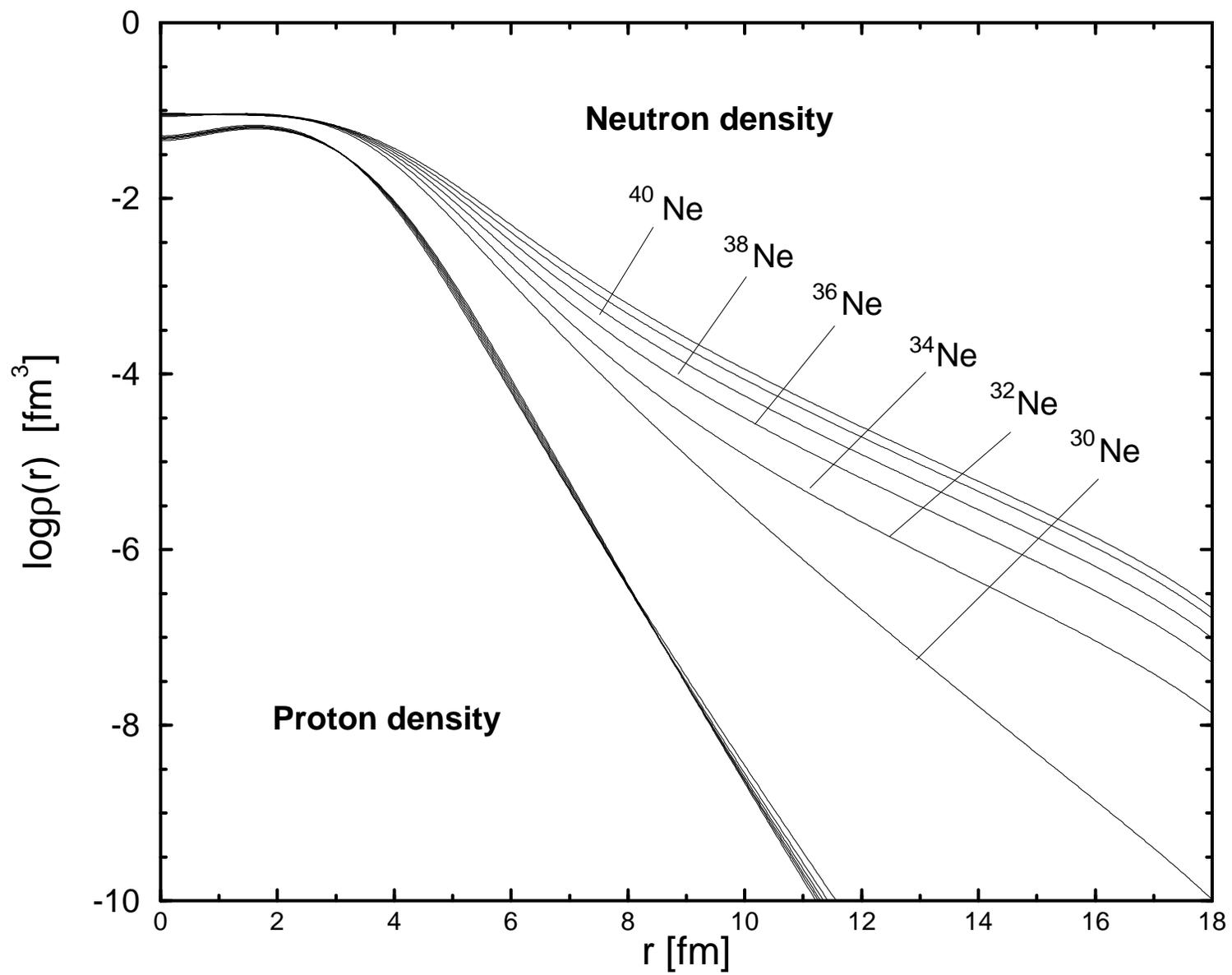
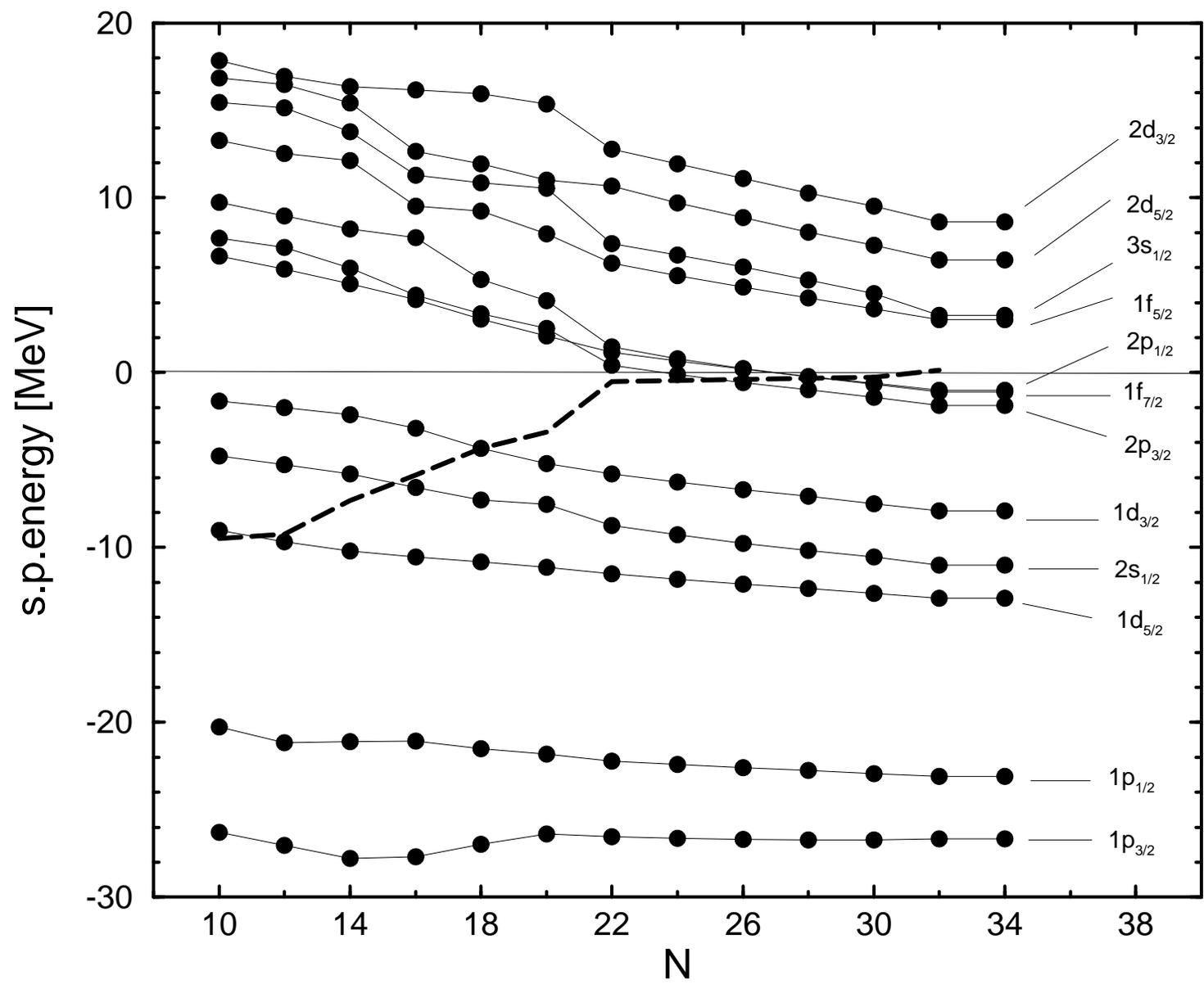
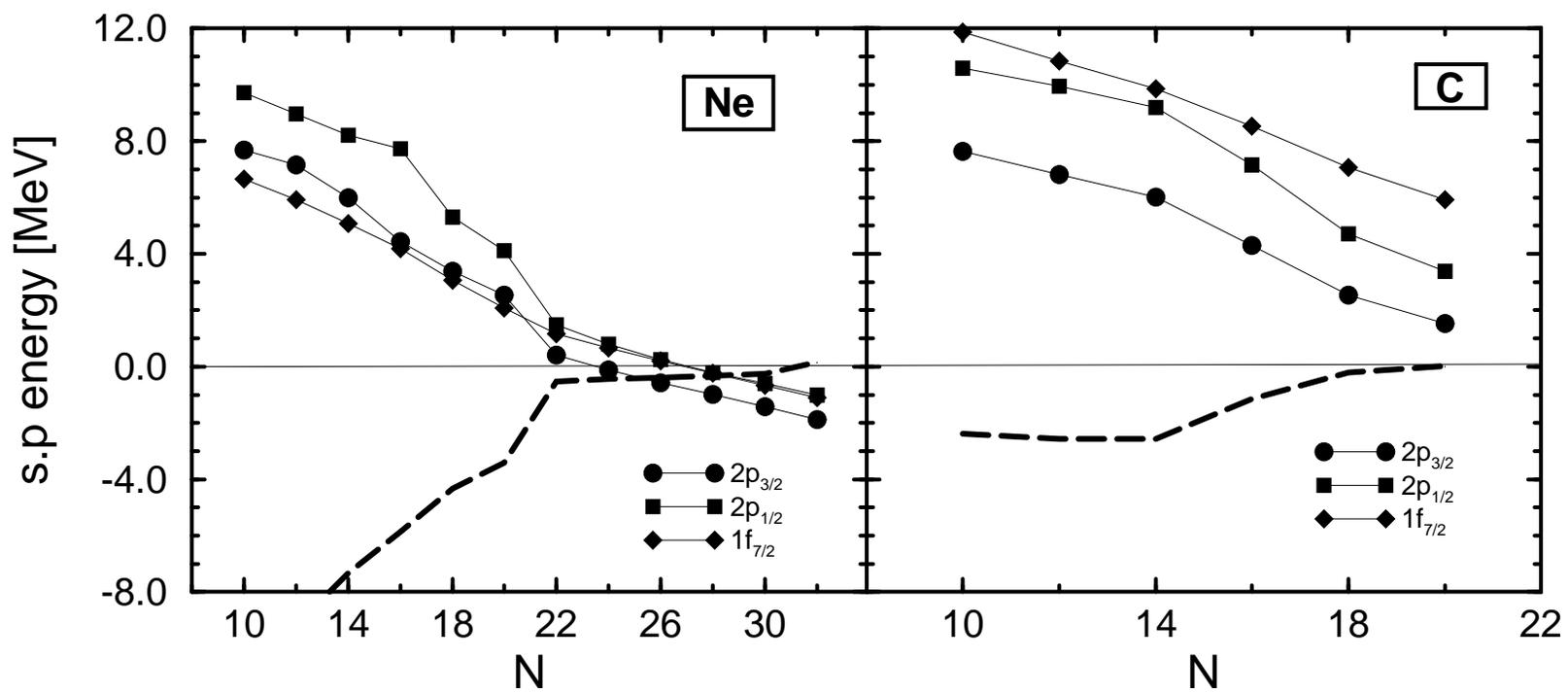
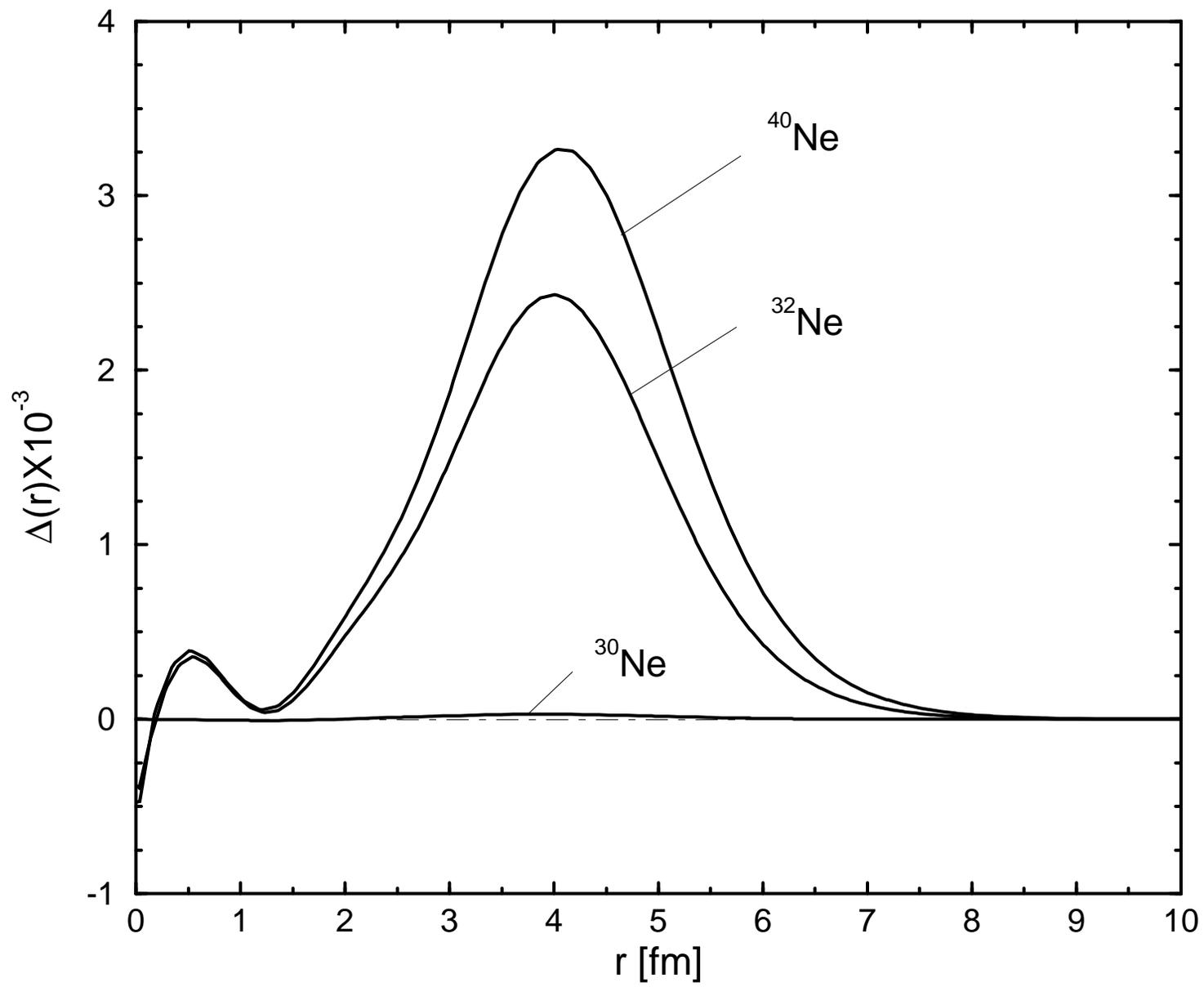
\end{document}